# Uterine muscle networks: Connectivity analysis of the EHG during pregnancy and Labor

**Noujoud Nader[1,2], Mahmoud Hassan[3], Wassim Falou[1,4], Mohamad Khalil[1,4],**

**Brynjar Karlsson[5], Catherine Marque[2]**

(1): Azm Center in Biotechnology and its Application, Lebanese University, Tripoli
(2): Sorbonne Universities, University of Technology of Compiegne, CNRS, UMR 7338 BMBI, Compiegne, France C
(3): LTSI, Université de Rennes 1, F-35000, France
(4): CRSI, Engineering Faculty, Lebanese university, Lebanon
(5): Reykjavik University, School of Science and Engineering, Reykjavik 101, Iceland

**Correspondence to:** Noujoud Nader, nader.noujoude@hotmail.com

ABSTRACT

In this paper, we propose a new framework to analyze the electrical activity of the uterus recorded by electrohysterography (EHG), from abdominal electrodes (a grid of 4x4 electrodes) during pregnancy and labor. We evaluate the potential use of the synchronization between EHG signals in characterizing electrical activity of the uterus during pregnancy and labor. The complete processing pipeline consists of i) estimating the correlation between the different EHG signals, ii) quantifying the connectivity matrices using graph theory-based analysis and iii) testing the clinical impact of network measures in pregnancy monitoring and labor detection.

We first compared several connectivity methods to compute the adjacency matrix represented as a graph of a set of nodes (electrodes) connected by edges (connectivity values). We then evaluated the performance of different graph measures in the classification of pregnancy and labor contractions (number of women=35). A comparison with the already existing parameters used in the state of the art of labor detection and preterm labor prediction was also performed. Results show higher performance of connectivity methods when combined with network measures. Denser graphs were observed during labor than during pregnancy. The network-based metrics showed the highest classification rate when compared to already existing features. This network-based approach can be used not only to characterize the propagation of the uterine contractions, but also may have high clinical impact in labor detection and likely in the prediction of premature labor.

# 1. INTRODUCTION

Worldwide an estimated 11.1% of all live births in 2010 were born preterm, with increasing rates in preterm birth in most countries [1]. Serious complications of preterm birth account for one million deaths each year, and preterm birth is the main risk factor in over 50% of all neonatal deaths. The immediate neonatal intensive care incurs large economic costs of preterm birth, including long-term complex health needs [1]. For this reason, many studies focused on pregnancy monitoring techniques to assess the key risk factors and allow the prediction of preterm labor. Labor is well-known to be preceded by two physiological phenomena: increased excitability and increased interactions between the myometrial cells which results in an increase of the propagation of the action potential that underlie uterine contractions [2]. It indeed permits to access the uterine excitability (with only one - signal, monovariate approach) as well as the synchronization of the uterine activity, by using multiple signals (bivariate approach).

Two notions of synchronization can be distinguished i) *local* where spike-level analysis is performed [3] and ii) *global* where the electrical activity is propagated over whole uterine burst due to electro-mechanical couplings phenomena [4]. Many other studies have reported that the analysis of the electrical uterine activity, recorded in a noninvasive way on the mother's abdomen, called the electrohysterography (EHG), can be a very powerful tool to monitor pregnancy and predict preterm labor [5] [6]. Other studies used the nonlinear correlation coefficient to estimate the relationships between 12 bipolar EHG derived from a matrix of 4x4 electrodes placed on the woman's abdomen [7][8]. In these studies, the nonlinear correlation coefficient was applied on the entire uterine burst, manually segmented. Authors showed a significant difference between pregnancy and labor contractions [7] as well as an increase in the correlation of EHGs as labor approaches [8]. A recent study by Govindan et al. [9] showed also an increase in the synchronization of the uterine activity few days before labor, using Magnetomyography (MMG).

Propagation velocity (*PV*) and conduction velocity (*CV*) have been also used to predict preterm labor [10], [11]. *CV* was quantified by analyzing either the propagation of whole bursts of EHG [10] [12], or single spikes identified within bursts [10][3][13][14]. The combination of *PV* (computed from spikes) and peak frequency (*PF*) reported so far the highest (96%) performance to discriminate labor and nonlabor contractions [10]. The analysis based on spikes (often by using small and close electrodes) would permit to quantify the electrical diffusion process. The one made from whole bursts (with larger and more spaced electrodes) would focus more on the global synchronization of the uterus.

Concerning the global analysis (whole burst), in most previous studies, the EHG connectivity matrices were reduced by keeping only their mean and standard deviations. Despite the encouraging results obtained, relevant information was missed due to this averaging. To better quantify the connectivity matrices, we proposed here the use of network measures (from graph theory-based analysis) combined with its clinical impacts in pregnancy monitoring and labor detection. This field has shown a growing interest in the last decade, especially to characterize brain networks [15]–[17]. According to this approach, the correlation matrix can be represented as graph of a set of nodes (electrodes) interconnected by edges (connectivity values between electrodes). In addition, most previous EHG connectivity-based studies focused on small database, which limited the possible conclusions from clinical viewpoint. Here, we have included, to our knowledge, the largest number of 16 channels EHG recordings so far (35 women, 430 contractions).

A high variability among connectivity methods in the context of EHG analysis were recently reported that could be likely related to the field spread effect when realizing connectivity analysis at electrodes level. The imaginary part of the coherence (*Icoh*), proposed in [18], was shown to reduce this effect (in the context of brain connectivity). In this paper, we show the clinical impact of graph theory based analysis to connectivity matrices (computed using *Icoh*) obtained from EHG signals recorded from women during pregnancy and/or labor. Labor detection and pregnancy monitoring using this approach is presented as well as node-wise analysis to reveal nodes (channels) the most discriminators.

## 2. MATERIALS AND METHODS

### 2.1. Overview

The complete pipeline of our approach is presented in Figure 1. The first step consists of recording the EHG signals using a grid of 4x4 monopolar electrodes (Figure 1a). Two reference electrodes were placed on each of the woman's hip. The EHG signals were then segmented, based on the tocodyanmometer signals reflecting the mechanical activity of the abdomen, and denoised (Figure 1b). The segmentation was used to isolate the EHG burst related to a contraction, evidenced on the Toco, from the non-contractile parts of the signal (baseline).

The third step is to compute the statistical couplings between the denoised signals by using connectivity measures (Figure 1c). The obtained connectivity matrix can be represented as a graph (Figure 1d). These graphs are computed for pregnancy and labor contractions at different term (Figure 1e). Several measures can be extracted from the obtained graphs based on graph theory (Figure 1f). These measures will be used to evaluate the clinical impact of the proposed approach in the classification of pregnancy and labor contractions and for pregnancy monitoring (Figure 1g), in term of global synchronization analysis.

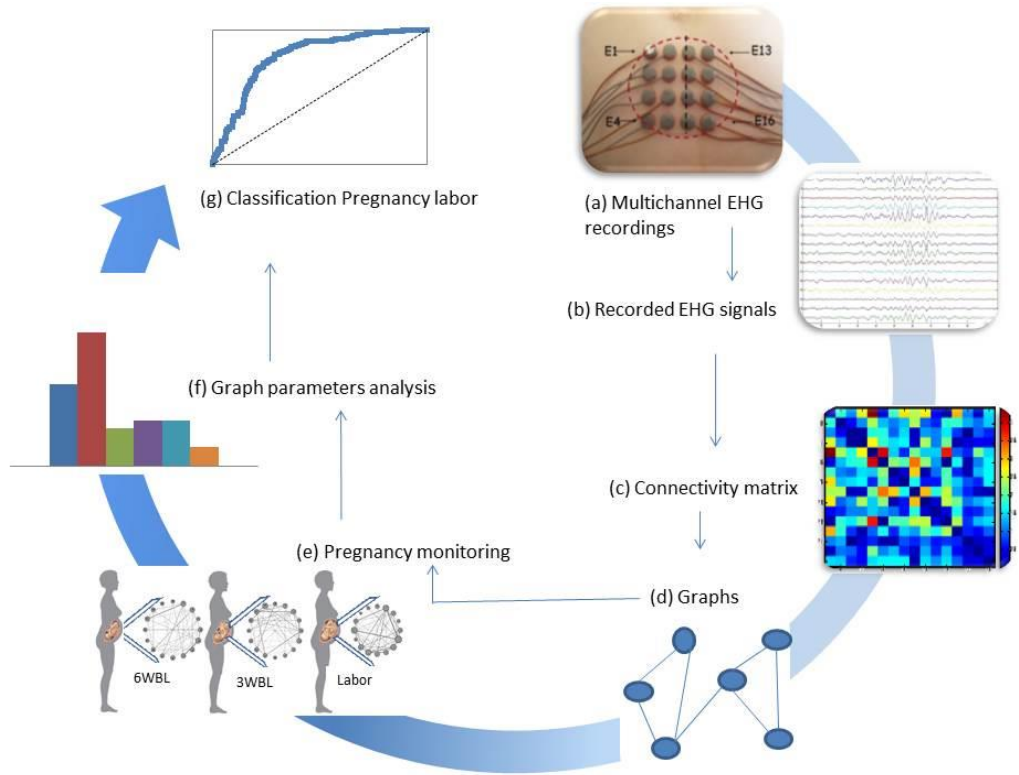

**Figure 1.** *Structure of the investigation.* (a) Multichannel EHG recordings using a grid of 4x4 electrodes. (b) Segmentation and filtering of EHG signals. (c) Pair-wise connectivity matrix. (d) Characterization of connectivity matrices using network measures (e) Graphs used for pregnancy monitoring along week of gestation. (f) Statistical study based on the extraction of graph parameters. (g) Classification of labor/pregnancy.

## 2.2. Data

To record the electrical activity of the uterine muscle, a grid of 16 monopolar electrodes (4x4 matrix) was placed on the woman's abdominal skin. The third electrode column was always placed on the uterine median vertical axis and the $10^{th}$–$11^{th}$ electrode midway between the symphysis and the uterine fundus. The reference electrodes were positioned on each of the woman's hips. The hip was chosen as reference as there is little electrical activity under the electrode and the distance from the reference electrodes to the abdominal electrodes is suitable. The recording system has an anti-aliasing filter with a high cut-off frequency of 100 Hz. The sampling frequency was 200 Hz. The data were recorded at the Landspitali university hospital (Reykjavik, Iceland) using a protocol agreed by the relevant ethical committee (VSN02-0006-V2) and at the Center for Obstetrics and Gynecology (Amiens, France), using a protocol agreed by the relevant ethical committee (ID-RCB 2011-A00500-41). A part of this database is already available on Physionet (the Icelandic data) [19]. The typical duration of a pregnancy recording was one hour while the duration of a labor recording was at about half an hour (depending on the delivery condition).

The tocodynamometer (Toco) paper trace was digitalized and used as reference in the EHG segmentation process. The EHG bursts related to uterine contractions (mechanical activity) were segmented manually based on the Toco. In order to deal with the different artifacts such as the fetal/mother cardiac activity, electronic noise and pump noise (appeared during labor), these segmented data were then denoised using CCA-EMD method, developed in the team [20]. After segmentation and denoising, we obtained 183 labor and 247 pregnancy bursts. Contractions were considered in labor if they are recorded with a delay of less than 24 hours before delivery. The average length of the contractions in pregnancy was 1 minute and in labor was 45 second. These contractions were extracted from 35 women. The analysis described below has been applied to these segmented contractions. Detailed information on the women included in this study is presented in Annex I. As inclusion criterion, woman should only have a monofetal pregnancy, thus women with twins were excluded from the study. In addition, women with any pathological issues were excluded. The information in Annex I include the weight (kg), height (m), week of pregnancy (WP; the day when data were recorded), in which group woman is enrolled (week before labor or labor group), week of delivery (WD) and the number of segmented contractions.

## 2.3. Connectivity Methods

In this paper, we compare three connectivity measures (see [21] for review about connectivity measures). The first method is the linear correlation coefficient ($R^2$). The second method is a modified version of the nonlinear correlation coefficient *(FW-$h^2$)* proposed in [22]. This method was chosen as it showed the highest performance for uterine EHG analysis so far. The third method is the imaginary part of the coherence (*Icoh*) proposed by [18]. This method was chosen for its robustness to the volume conduction problem, a problem supposed to dramatically affect the correlation at the electrode level (as shown in the context of brain connectivity for instance). More details about the three methods are described below:

## The Linear Correlation ($R^2$)

The linear correlation coefficient between two time series $X(t)$ and $Y(t)$, in the time domain, is defined as:

$$R^2 = \max_{\tau} \frac{cov^2(X(t), Y(t+\tau))}{var(X(t))var(Y(t+\tau))} \qquad (1)$$

where *var* and *cov* denote respectively variance and covariance between $X(t)$ and $Y(t)$. $\tau$ denotes the time delay [21].

## Filtered-Windowed- $h^2$ (FW-$h^2$)

The nonlinear correlation coefficient ($h^2$) between two signals $X(t)$ and $Y(t)$ of length N is defined as:

$$h^2_{Y/X} = \frac{\sum_{k-1}^{N} Y(k)^2 - \sum_{k-1}^{N}(Y(K) - f(X_i))^2}{\sum_{k-1}^{N} Y(k)^2} \qquad (2)$$

Where $f(X_i)$ is the linear piecewise approximation of the nonlinear regression curve. $h^2_{Y/X}$ value varies between 0 (Y and X are independent) and 1 (Y and X are fully correlated) and $h^2_{Y/X} \neq h^2_{X/Y}$ [21], [23]. Here, we used the modified version of $h^2$ proposed by Diab et al [22]. This method, called Filtered-Windowed- $h^2$ (FW_$h^2$), has showed highest performance in labor detection when compared to other classical methods. Briefly, a preprocessing procedure was introduced before computing $h^2$. It consist of filtering the data in very low frequency bands (0.1 – 0.3 Hz) and then segmenting the uterine contractions using the bivariate piecewise stationary signal pre-segmentation (bPSP) algorithm proposed in [24].

## The Imaginary part of Coherence (Icoh)

One of the problem faced when computing correlation at the surface level is the so called 'volume conduction' problem. This term is usually used to describe the tissues layered between the source of a signal (the muscle in our case) and the recording site (the woman abdomen). This volume conductor is known to affect the electrical activity of the uterine muscle, when recorded at the skin surface [25]. Thus, a correlation between several electrodes can be detected even if the signals come from the same sources due to the diffusion of signals across the volume conductor.

To deal with this problem, the imaginary part of the coherence (*Icoh*) method was proposed by Nolte et al [18]. The hypothesis behind this method is that the real part of the coherence function reflects the zero lag interactions between signals and thus the imaginary part of the coherence may reflects the true (unbiased) interactions [18]. The coherence function reflects the linear frequency-dependent correlation between *X* and *Y*. After estimating the cross-spectral density function $C_{XY}$ of the two signals, a normalization step can be performed using the individual auto-spectral density functions $C_{XX}$ and $C_{YY}$. The *Icoh* is then defined as:

$$Icoh = \frac{|Im C_{XY}(f)|}{\sqrt{|C_{XX}(f)||C_{YY}(f)|}} \tag{3}$$

### 2.4. Graph Theory

The connectivity matrix, calculated between all possible electrode pairs, is then represented as a graph. A graph is an abstract representation of a network, consisting of a set of nodes (*N*) connected by edges (*V*) [26]. In our case, the nodes represent the electrodes (N=16) and the edges represent the value of the connectivity measure. Here, we extract two different graphs metrics:

**Strength**

The strength shows the importance and the contribution of each node with respect to the rest of the network and defined as:

$$S_i = \sum_{j \in N} w_{ij}. \tag{4}$$

where *i*, *j* denotes respectively the i[th] and j[th] nodes and $w_{ij}$ is the value (weight) of the connectivity between nodes *i* and *j* [15]. The average strength value over all the nodes can be also computed, reflecting the overall characteristic of the network.

### 2.5. Existing Methods

**Propagation Velocity and Peak Frequency (PV+PF)**

Lucovnik et al. have explored the performance of the Propagation Velocity (*PV*) in the discrimination of nonlabor and labor EHGs [10]. After estimating the distance *d* that the propagating signals travels and the time *t* needed for crossing this distance, *PV* can be estimated by dividing the distance *d* by the time *t*. For a given EHG, after computing the Peak Frequency (*PF*) from its power spectral density, the obtained *PF* value is then combined with its *PV* values by a simple addition of the two metrics.

**Conduction Velocity (CV)**

The Conduction Velocity (*CV*) was proposed by Rabotti et al. [3], [11]. Authors estimated the velocity and the direction of the propagation of different spikes identified from the EHG signals. The time delay between two electrodes at a given row is $t_r$ and at a given column is $t_c$. The velocity *v* and the angle of propagation *Θ* were computed as following:

$$v = f_s \frac{d\cos(\theta)}{\tau_r}$$
$$v = f_s \frac{d\sin(\theta)}{\tau_c} \qquad (5)$$

Where $f_s$ is the sampling frequency. For more details, see [3].

In our study, CV and PV were applied to the entire burst and not the single spikes, as originally proposed. This choice was made to standardize the comparison with the correlation-based methods.

**2.6. Data analysis**

A total number of 247 pregnancy and 183 labor contractions were segmented from 35 women. In order to differentiate between these two groups, we have computed three connectivity methods: $R^2$, *FW_h²* and *Icoh*. As all these methods are bivariate, the connectivity measure was computed between each pair of the 16 signals. For each channel the connectivity measure was computed with all other channels providing 16×16 values ranging between 0 (no connectivity) and 1 (strongly connected). These 16x16 values were represented in form of matrix where the connectivity values were encoded by color from blue (weak connectivity) to red (strong connectivity). Thus, we obtained a connectivity matrix (16×16) for each contraction and each method. We have then tested the performance of each method for the classification of pregnancy and labor contractions. To investigate the added value of the graph theory based analysis, we have compared the results given by each graph metric with the ones obtained by the approach previously used (average of the weight values of each connectivity matrix). The results were also compared to *PV+PF* and *CV* using ROC curves.

To explore the evolution of the uterine muscle networks from pregnancy to labor, we have categorized the uterine contractions in 11 groups of weeks before labor (WBL) effecting the number of weeks before delivery (date of which is known for every woman included in our database), in addition to one labor group. Some of these groups contained contractions from only one woman. For this reason, only the WBL groups that contain more than 4 women and more than 18 contractions per group were used here to avoid, as possible, the unbalanced number of women/contractions between the classes. As result, we kept the following groups: 8 WBL, 6 WBL, 4 WBL, 3 WBL, 2 WBL, 1 WBL and labor.

### 2.7. Statistical Tests

Wilcoxon test was used to test the significant differences obtained between conditions. To evaluate the classification performance of the different features, we used the Receiver Operating Characteristic (ROC) curve analysis. ROC curve is a standard tool for diagnostic classification test evaluation. In a ROC curve the true positive rate (sensitivity) is plotted in function of the false positive rate (1-specificity) for different cut-off points of a parameter. The area under the ROC curve (AUC) is a measure of how well a parameter can distinguish between two diagnostic groups (in our case labor vs. pregnancy). We have also used as the regression line that indicates the general tendency of the strength values points. We used the Brain Connectivity Toolbox (BCT) for the calculation of graph parameters[15]. For the graph visualization, we used the 'GEPHI' software [27].

## 3. RESULTS

### 3.1. Labor Detection

Figure 2 presents the ROC curves obtained for each tested method. The AUC was higher when using the graph parameters for *Icoh* (AUC increases from 0.504 to 0.801, Figure 2b) and *FW-h$^2$* (AUC increases from 0.658 to 0.77, Figure 2a). The results obtained when using only $R^2$ (AUC=0.669) are very close to the ones obtained when using graph measures (AUC=0.664), Figure 2c. The AUC obtained with the *CV* was 0.495 while for *PV+PF* was 0.789.

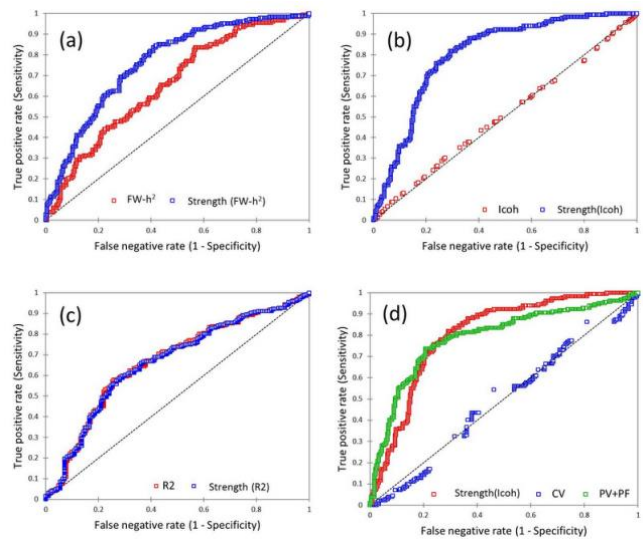

**Figure 2.** ROC curves for methods without and with using graph analysis (strength parameter). (a) ROC curve for *FW_h$^2$* and strength (*FW_h$^2$*). (b) Roc curve for *Icoh* and strength(*Icoh*). (c) ROC curve for *R$^2$* and strength(*R$^2$*). (d) Comparison of *CV*, *PV+PF* and Strength(*Icoh*).

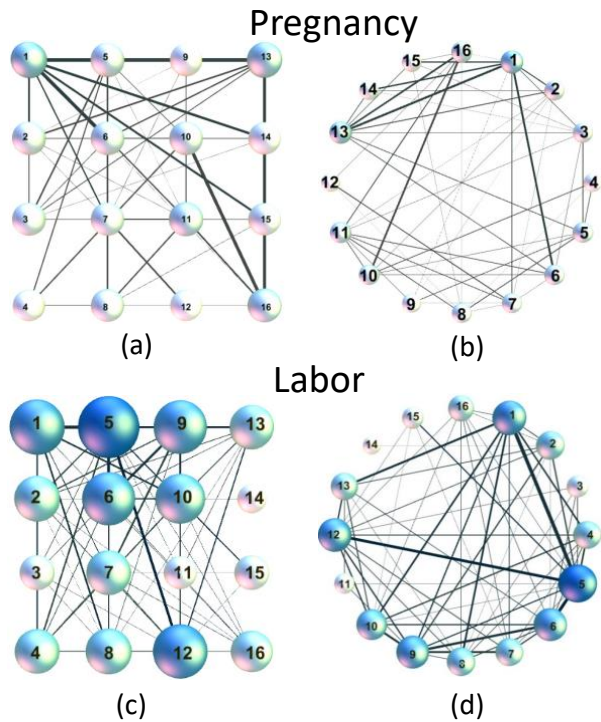

**Figure 3.** Graph results using *Icoh* (a-b) Mean pregnancy graph (d-c) Mean labor graph

Figure 3 shows the graphs averaged over the 247 pregnancy (a-b) and the 183 labor (c-d) contractions using *Icoh*.

In Figures 3a and 3c, we represent each graph as a grid of 4x4 nodes (electrodes) as located on the woman's abdomen. The edges represent the connectivity values between electrodes. In Figures 3b and 3d, we present the same results in a circular layout. The thickness of each edge depends on its weight (here *Icoh* values). The size and the color of a node depend on its strength value. Figure 3c and 3d showed that the nodes 1, 5 and 12 have the highest strength values and that the weights are the highest (thickest edges) between nodes 1-5 and also between nodes 5-12 for the labor graph.

We then computed the statistical test at each electrode. Boxplots are shown Figure 4. The figure shows an increase in the strength values from pregnancy to labor at all electrodes. Indeed, for all the electrodes the boxplots were higher in labor compared to pregnancy. This difference was significant for all the electrodes ($p<0.01$, corrected for multiple comparisons using Bonferroni). The highest significance was observed at node 1, 5 and 12.

### 3.2. Pregnancy Monitoring

In this section, the performance of the proposed approach for the monitoring of pregnancy evolution along term is presented. Figure 5a shows the evolution of the average strength values for each woman at each WBL. Values were slightly changing from 8WBL to 1WBL, while an increase between 1WBL and labor groups was observed. Figure 5b-h shows also the corresponding averaged graphs for the different terms. We can notice that the number of significant edges in the averaged labor graph (Figure 5h, density=0.5) was higher than the different terms (density was always lower than 0.4). In terms of nodes diameter, no significant changes were obtained before labor, while a difference for nodes 1, 5 and 12 can be observed when going from 1WBL to labor.

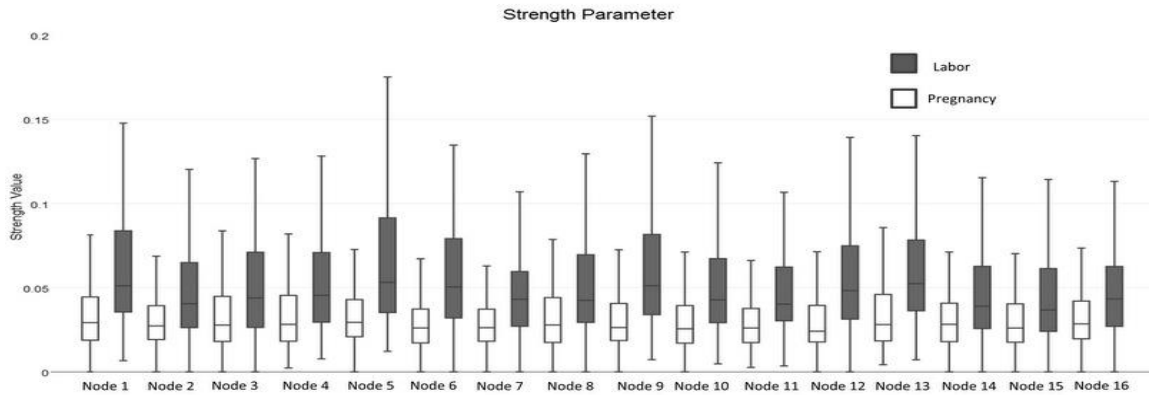

**Figure 4.** Boxplots of strength values in pregnancy and labor on 16 nodes (electrodes). All the differences are significant ($p<0.01$).

We have then computed the value of Strength (*Icoh*) at each node for pregnancy monitoring. In figure 6 we have computed the value of Strength (*Icoh*) for node 12 for the monitoring of pregnancy evolution along term and we have presented the boxplot for each term. In addition to its location on the median axis, the node 12 was selected as it showed the highest classification rate. Figure 6a shows that all the strength values during pregnancy were relatively low, with increased values during labor. We present in Figure 6b-h the corresponding averaged graphs for each of the term groups. We highlighted in each graph only node 12 and the nodes to which it connects. We can observe in the labor graph that node 12 is associated to a high number (11/15) of significant edges (Figure 6h) in contrast to pregnancy, where node 12 connects to a maximum of 6 nodes at almost all the pregnancy groups (Figure 6b-g). In terms of node strength, no significant difference was observed between all the WBL graphs, while during labor node 12 was larger (higher strength) than pregnancy groups. No significant difference was observed between the pregnancy groups, except between 8WBL and 2WBL ($p=0.009$) while a significant difference was obtained between labor and all the other groups ($p<0.01$, corrected for multiple comparison using Bonferroni).

To guarantee that this network reconfiguration is actually related to the labor process and not only resulted from the simple progressing of the gestation, we selected contractions from women recorded at the same term, 39 WG (weeks of gestation, counted from the time of their last menstrual period) but some being already in labor (labor group) and the others having delivered later (pregnancy group). We have 11 contractions from 5 women in the pregnancy group, and 41 contractions from 5 women in labor. We present in Figure 7 the difference in the connectivity networks for these two groups recorded at 39WG. A qualitative difference is observed between the mean graph of pregnancy (Figure 7a) and the mean graph of labor (Figure 7b) in term of edges weight and node strengths. All these values were higher in the labor group.

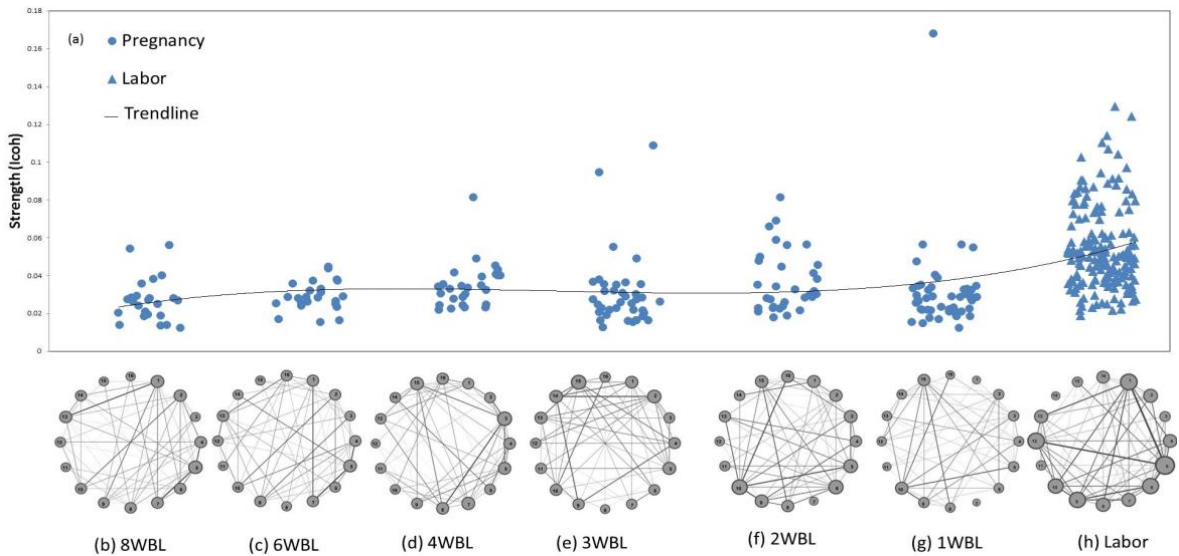

**Figure 5**: (a) Evolution of Strength (*Icoh*) with week before labor. Each point represents the strength value of one contraction for a given woman. Mean graph for: (b) 8WBL. (c) 6WBL. (d) 4WBL. (e) 3WBL. (f) 2WBL. (g) 1WBL. (h) Labor.

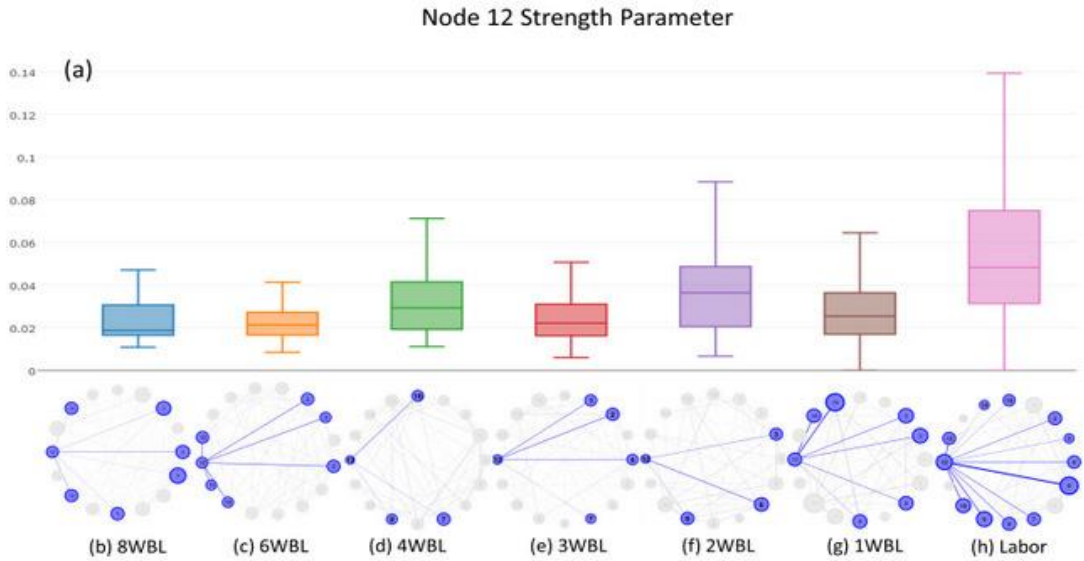

**Figure 6.** Boxplot of strength values for node 12 from with week before labor. Mean graph for: (b) 8WBL. (c) 6WBL. (d) 4WBL. (e) 3WBL. (f) 2WBL. (g) 1WBL. (h) Labor.

## 4. DISCUSSION AND CONCLUSION

In this paper, we have presented the results of a novel approach aiming at characterizing the functional connectivity of the uterine electrical activity. We investigated the power of the network-based analysis to characterize the evolution of uterine contractions from pregnancy to labor and to discriminate pregnancy and labor contractions. Previously, the connectivity matrices computed from EHGs were usually transformed to a single value per contraction, by averaging the connections weights of each matrix [7]. Consequently, useful information was certainly lost. Indeed, the graph theory-based analysis used here seems to be a better way to quantify the connectivity matrix.

In this study, the graph theory-based analysis has been proven to be more efficient to quantify connectivity matrices for normal pregnancy and labor contractions than the previous classical quantification of the connectivity matrices. However, the method showed lower performance for pregnancy monitoring as no significant changes were observed between the different pregnancy weeks before labor. These results are more specifically discussed hereafter.

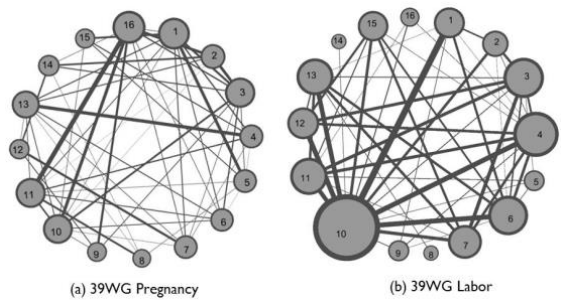

**Figure 7.** Mean graphs for EHGs recorded at 39WG: (a) Pregnancy, (b) Labor.

**Increase of synchronization with term**

One of the key results obtained in this work using graph theory -at the whole burst level - is an increase in synchronization with increasing pregnancy term. This finding agrees with the previously reported results using EHG [7] or MMG-based studies where authors showed an increase in synchrony as the women approach active labor [9].

This network-based approach has improved the classification between pregnancy and labor. The results obtained with Strength(*Icoh*) (AUC=0.801) were higher than those obtained by *PF+PV* (AUC = 0.789), as well as by *CV* (AUC=0.495). It is however difficult to compare these results with the reported good performance of *PV/CV* in previous analysis [10][11], [14] as these metrics were computed differently. *PV* and *CV* were usually applied to single spikes not to whole uterine burst which may explain the reported poor results of both methods in our study. We have computed *PV* and *CV* on the whole burst to standardize the computation way and to be able to compare with the correlation-based methods. These poor results of *CV* and *PV* do not put any doubt about their high performance of *CV* and *PV* when used on single spikes as reported in [3]. Nevertheless, with the whole burst approach, getting free from spike identification may present a huge advantage from the applicative point of view. In addition, a classification rate of 80% between labor and nonlabor groups is still not clinically sufficient. A possible improvement of these results can be the use of the EHG source connectivity approach (as realized recently in the context of brain connectivity) [28] and the possible combination of different features related to different physiological phenomena (excitability and propagation).

Results showed also an increased connectivity during labor. This increase in the strength values from pregnancy to labor was noticeable for all the electrodes. These findings are in agreement with the results obtained previously by Hassan et al. when using the nonlinear correlation coefficient on a smaller dataset [7]. A possible explanation of this increase in connectivity during labor is the propagation phenomenon, associated with the appearance of a large number of gap junctions prior to labor [29], as well as the electromechanical coupling proposed by Young as one of the synchronization process appearing during labor [4].

It is important to notice that all women included in our study gave birth at term (none of the births was premature). Our study showed the possible use of a new promising approach to first characterize the uterine bursts during pregnancy and labor and secondly, to classify normal pregnancy and labor contractions. To validate the clinical impact of the approach, the method will be applied to data from women with premature labors. In addition, different steps in our pipeline, such as the manual burst segmentation also should be automatized when going to the practical clinical use of this approach.

**Limitations**

First, a classical and still unsolved question relates to the setting of threshold values applied to the connectivity matrices. In this study, the same threshold value was used for each method, WBL or WG to standardize the analysis (10% of the maximum connectivity values). Other threshold values were also investigated (10% to 50%) and gave very similar results in term of differences between methods and conditions. Other approaches can be also explored like those based on surrogate data, although requiring a high computation time. Second, another unsolved question that presents a limitation for this study is low number of longitudinal recordings. Only few women (14) have been recorded at several weeks of gestation. Recording contractions during pregnancy is difficult, the women being available only when present at the hospital (for standard follow up, or hospitalization for risk pregnancy), and the contractions number is relatively low during pregnancy.

Third, it is important to keep in mind that the estimation of the functional connectivity at the electrode (surface abdomen) level can be affected by the volume conduction problem. The volume conduction is related to the fact that different channels are actually measuring the activity of a single uterine source. To tackle this problem, the imaginary part of the coherence function was used here as it was proven to have a high performance to reduce this effect in the context of brain connectivity [18]. Moreover, in the context of electroencephalography, the connectivity analysis at the brain source level showed a considerable reduction of the effect of the volume conduction when compared to the scalp level [30]. One possible improvement to the results reported in this study is to adapt the 'source connectivity' approach to the uterine muscle, by localizing the EHG sources at the uterine muscle level.

Finally, different methodological issues raised in the proposed pipeline such as the choice of the functional connectivity method or the graph metric. Therefore, the main focus of this paper was to evaluate those parameters in order to reveal an optimal combination that lead to the highest performance in term of classification of pregnancy vs. labor (normal) contractions. The next step is the application of this new approach to data recorded during pregnancy and labor for women with preterm birth.

To sum up, we showed that the network-based approach could be successfully used to not only characterize uterine electrical activity during pregnancy and labor but also classify pregnancy and labor contractions. This new approach could have a high clinical impact for detecting alterations in networks in relation with the abnormal (pathological) contractions associated to premature birth.

# ANNEX I

| Woman | Weight (Kg) | Height (m) | Week of pregnancy (WP) | Week of Delivery (WD) | Group | Number of contractions |
|---|---|---|---|---|---|---|
| W1 | 89 | 1.7 | 42 | 42 | Labor | 22 |
| W2 | 92.4 | 1.78 | 35 | 40 | 5 WBL | 5 |
| | | | 37 | | 3 WBL | 5 |
| | | | 38 | | 2 WBL | 6 |
| | | | 39 | | 1 WBL | 2 |
| W3 | 105 | 1.72 | 33 | 38 | 5 WBL | 1 |
| | | | 36 | | 2 WBL | 5 |
| | | | 37 | | 1 WBL | 3 |
| | | | 38 | | Labor | 10 |
| W4 | 67 | 1.64 | 34 | 38 | 4 WBL | 6 |
| | | | 36 | | 2 WBL | 7 |
| | | | 37 | | 1 WBL | 9 |
| W5 | 76.2 | 1.7 | 37 | 37 | Labor | 5 |
| W6 | 71 | 1.75 | 33 | 41 | 9 WBL | 7 |
| | | | 37 | | 4 WBL | 3 |
| W7 | 61 | 1.75 | 35 | 40 | 5 WBL | 7 |
| | | | 38 | | 2 WBL | 5 |
| | | | 39 | | 1 WBL | 6 |
| W8 | 62 | 1.65 | 33 | 39 | 6 WBL | 4 |
| W9 | 48 - 50 | 1.6 | 29 | 41 | 12 WBL | 2 |
| | | | 31 | | 10 WBL | 2 |
| | | | 34 | | 7 WBL | 1 |
| W10 | 75 | 1.72 | 36 | 40 | 4 WBL | 2 |
| | | | 38 | | 2 WBL | 3 |
| | | | 40 | | Labor | 1 |
| W11 | 70 - 75 | 1.76 | 33 | 41 | 8 WBL | 4 |
| | | | 35 | | 6 WBL | 2 |
| | | | 38 | | 3 WBL | 4 |
| W12 | 63.4 | 1.63 | 39 | 39 | Labor | 7 |
| W13 | 56 | 1.63 | 40 | 41 | 1 WBL | 8 |
| W14 | 100 | 1.78 | 33 | 41 | 8 WBL | 7 |
| W15 | 62 | 1.63 | 39 | 39 | Labor | 4 |
| W16 | 109 | xxx | 40 | 40 | Labor | 3 |
| W17 | xxx | xxx | 40 | 40 | Labor | 26 |
| W18 | xxx | xxx | 40 | 40 | Labor | 33 |
| W19 | xxx | xxx | 39 | 39 | Labor | 23 |
| W20 | xxx | xxx | 42 | 42 | Labor | 11 |
| W21 | xxx | xxx | xxx | xxx | Labor | 18 |
| W22 | 95 | 1.63 | 39 | 39 | Labor | 1 |
| W23 | 83 | 1.7 | 34 | 40 | 6 WBL | 1 |
| | | | 36 | | 4 WBL | 2 |
| | | | 37 | | 3 WBL | 4 |
| | | | 39 | | 1 WBL | 4 |
| W24 | 68 | 1.68 | 33 | 39 | 6 WBL | 7 |
| W25 | 69.5 | 1.67 | 31 | 39 | 8 WBL | 3 |
| | | | 36 | | 6 WBL | 4 |
| | | | 39 | | Labor | 4 |
| W26 | 95.3 | 1.62 | 34 | 39 | 5 WBL | 1 |
| W27 | 110 | 1.76 | 37 | 41 | 4 WBL | 1 |
| | | | 38 | | 3 WBL | 1 |
| | | | 39 | | 2 WBL | 2 |
| | | | 40 | | 1 WBL | 9 |
| W28 | 90 | 1.68 | 37 | 39 | 2 WBL | 1 |
| W29 | 85.5 | 1.68 | 32 | 40 | 8 WBL | 10 |
| | | | 37 | | 3 WBL | 9 |
| | | | 38 | | 2 WBL | 2 |
| W30 | 78 | 1.63 | 39 | 42 | 3 WBL | 1 |
| W31 | 113.3 | 1.73 | 36 | 39 | 3 WBL | 3 |
| W32 | 65.5 | 1.69 | 38 | 40 | 2 WBL | 6 |
| W33 | 74 | 1.68 | 37 | 41 | 4 WBL | 4 |
| W34 | 88 | 1.76 | 36 | 40 | 4 WBL | 1 |
| | 89 | | 39 | | 1 WBL | 1 |
| W35 | 82 | 1.67 | 33 | 40 | 7 WBL | 11 |
| | 83 | | 36 | | 4 WBL | 7 |
| | 84 | | 37 | | 3 WBL | 9 |
| | 85 | | 39 | | 1 WBL | 9 |